# SPIN-VIBRATIONAL $1^+$ STATES IN $^{140}$Ce

E.GULIYEV[1], Ö.YAVAŞ[1], A.A.KULIEV[2,3]([*])

[1] Ankara Universty Faculty of Science Department of Eng. Physics, 06100 Tandogan, Ankara, Turkey
[2] Sakarya University, Faculty of Science and Art, Physics Department
54100 Serdivan, Adapazarı, Turkey
[3] Permanent address: Institute of Physcs, Azerbaican Academy of Sciences, Baku, Azerbaijan
(*) E-mail: kuliev@mail.sakarya.edu.tr

**Abstract.** - The properties of collective $I^\pi = 1^+$ states in double even semi-magic nucleus $^{140}$Ce are investigated in random phase approximation (RPA). The calculation of the $B(M1)$ strength, elastic γ-scattering cross section $\sigma_\gamma(\omega)$ and γ-decay width $\Gamma_\gamma(\omega_i)$ show that in $^{140}$Ce the isovector spin-spin interactions concentrate the main strength of M1 transition at energy 7.89 MeV where the isovector spin-flip magnetic resonance may appear. These predictions are consistent with the observed M1 strength in $^{140}$Ce which is seen to be concentrated within a range of about 0.8 MeV centered at an excitation of 7.95MeV.

During the last few years great success has been achieved in measurements of the spin-flip magnetic dipole resonance in nuclei [1]. These experiments show that in heavy spherical and deformed nuclei there exists a very broad M1 resonance at energies between 7 and 11 MeV centred at energy on the order of $44A^{-1/3}$ MeV. Since long ago some experimental evidence for unusual concentration of M1 strength in nucleus $^{140}$Ce at energy between 7 and 11 MeV has been obtained by inelastic (e,e') [2], (p,p') [3], (n,γ) [4] and tagged photon scattering [5]. Indeed because of the limited energy resolution no individual state could be discerned in the existence scattering experiments in the resonance energy region. Recent technical improvements of the photon scattering experiments [6] and a new method proposed to investigate the scattering reactions with photons on fully ionised nuclei, namely FEL γ-nucleus colliders [7,8] open the way for such studies up to excitation energies of about 10 MeV. Therefore the study of the magnetic dipole giant resonance in heavy nuclei has attractive considerably interest.

By now there are several papers dealing with spin-vibrational $1^+$ excitations in $^{140}$Ce [9,10]. In these papers, however, either rather schematic calculations are performed, or on pays special attention only to the collective $1^+$ state, which should appear at relatively high excitation energy.

In connection with the increasing interest of investigating the M1 transitions to highly-excited states, it is interesting to study in the microscopic approach the properties of spin-vibratıonal $1^+$ states generated by the isovector spin-spin interactions and estimate their energy, M1-strength and elastic photon scattering cross section in $^{140}$Ce.

In this paper collective $1^+$ states are treated as vibrational states in the framework of quasi-particle random phase approximation (QRPA). Supposing the collectivisation of states being due to an isovector spin-spin interactions [11], then model Hamiltonian of system can be written as

$$H = H_{sp} + H_{pair} + V_{\sigma\tau} \tag{1}$$

where

$$V_{\sigma\tau} = \frac{1}{2}\chi_{\sigma\tau}\sum_{i,j}\vec{\sigma}_i\vec{\sigma}_j\tau_i^z\tau_j^z \tag{2}$$

Here $H_{sp}$ is the hamiltonian of the single-particle motion in an average field and $H_{pair}$ is the pairing interaction hamiltonian. σ and τ are respectively, the spin and isospin matrices. The

isovector spin-spin interactions (2) can be written in terms of particle operators in the following form:

$$V_{\sigma\tau} = \frac{1}{2}\chi_{\sigma\tau}\sum_{\mu=0,\pm1}T_\mu^+ T_\mu \qquad (3)$$

where

$$T_\mu = \sum_{(jm)}\frac{\langle j\|\vec{\sigma}\|j'\rangle}{\sqrt{2j+1}}\langle j'm'1\mu|jm\rangle a_{jm}^+ a_{j'm'}.$$

Here $a^+(a)$ are creation (annihilation) operators for nucleon and $\langle j\|\vec{\sigma}\|j'\rangle$ are the single-particle reduced matrix elements of the spin operator σ [9]

In the quasi-particle representation the $T_\mu$ operator is of the form

$$T_\mu = \sum_{jj'}\sigma_{jj'}\{M_{jj'}D_{jj'}(\mu) + \frac{1}{2}L_{jj'}[C_{jj'}^+(-\mu) + (-1)^\mu(C_{jj'}(\mu))]\} \qquad (4)$$

Where $\sigma_{jj'} = \frac{1}{\sqrt{3}}\langle j\|\vec{\sigma}\|j'\rangle$ and the operators $C_{jj'}$ and $D_{jj'}$ are defined as

$$C_{jj'}(\mu) = \sqrt{\frac{3}{2j+1}}\sum_{mm'}(-1)^{j'-m'}\langle j'm'1\mu|jm\rangle\alpha_{j'-m'}\alpha_{jm}$$

$$D_{jj'}(\mu) = \sqrt{\frac{3}{2j'+1}}\sum_{mm'}\langle jm1\mu j'm'\rangle\alpha_{j'-m'}^+\alpha_{j-m} \qquad (5)$$

Here $\alpha^+(\alpha)$ are quasi-particle creation (annihilation) operators and $M_{jj'} = u_j u_{j'} + v_j v_{j'}$ and $L_{jj'} = u_j v_{j'} - u_{j'} v_j$ and $u_j$ ($v_j$) are the Bogolyubov quasi-particle transformation parameters.

In RPA the excited $1^+$ states are defined single-phonon states:

$$|\Psi_i(\mu)\rangle = Q_i^+(\mu)|\Psi_0\rangle = \frac{1}{\sqrt{2}}\sum_\nu\{\psi_\nu^i C_\nu^+(\mu) - (-1)^{\mu+1}\varphi_\nu^i C_\nu(-\mu)\}|\Psi_0\rangle \qquad (6)$$

where $|\Psi_0\rangle$ is the phonon vacuum–the ground states of even-even nuclei and $Q_i^+$ is $1^+$ phonon creation operator. In order to simplify the notation we used, instead of the pairs indices $(j,j')$, a single index ν. In further the sum $\Sigma^\tau$ denotes summation over the states of one kind of nucleons and Σ denotes the summation over all neutron and proton states. The two-quasi-particle amplitudes $\psi_\nu^i$ and $\varphi_\nu^i$ are normalised by

$$\sum_\nu(\psi_\nu^{i\,2} - \varphi_\nu^{i\,2}) = 1 \qquad (7)$$

Employing the conventional procedure of RPA and solving the equation of motion

$$[H_{sp} + H_{pair} + V_{\sigma\tau}, Q_i^+] = \omega_i Q_i^+ \qquad (8)$$

we obtain the dispersion equation for the excitation energy $\omega_i$ of $1^+$ states:

$$1 + \chi_{\sigma\tau}[F_n(\omega_i) + F_p(\omega_i)] = 0 \qquad (9)$$

where

$$F_\tau = \sum_v {}^{(\tau)} \frac{2\varepsilon_v \sigma_v^2 L_v^2}{\varepsilon_v^2 - \omega_i^2} \qquad (10)$$

Here $\varepsilon_v = \varepsilon_j + \varepsilon_{j'}$ are two quasi-particle energies. For the amplitudes $\psi_v^i$ and $\varphi_v^i$ of the single-phonon states with energy determined by the solution of (9), we have

$$\psi_v^i(q) = -\frac{q}{\sqrt{Y(\omega)}} \frac{\sigma_v L_v}{\varepsilon_v - \omega_i} \qquad (11)$$

$$\varphi_v^i(q) = -\frac{q}{\sqrt{Y(\omega_i)}} \frac{\sigma_v L_v}{\varepsilon_v + \omega_i} \qquad (12)$$

where

$$Y(\omega_i) = 4\omega_i \sum_v \frac{\varepsilon_v \sigma_v^2 L_v^2}{(\varepsilon_v^2 - \omega_i^2)^2} \qquad (13)$$

Here q=1 for neutron and q=−1 for proton amplitudes, respectively.

A more important characteristic of $1^+$ states is the reduced M1 transition probability, which in the RPA can be written in the form

$$B(M1, 0_g^+ \to 1_i^+) = \frac{3}{4\pi} \frac{1}{\sqrt{Y(\omega_i)}} [J_p(\omega_i) + \sum_\tau (g_s^\tau - g_l^\tau) F_\tau(\omega_i)] \mu_N^2 \qquad (14)$$

Where for one kind of nucleon

$$J_p(\omega_i) = \sum_v {}^{(p)} \frac{2\varepsilon_v j_v^2 L_v^2}{\varepsilon_v^2 - \omega_i^2} \qquad (15)$$

In eq.(14) $g_s$ and $g_l$ are spin and orbital gyromagnetic ratios of the free nucleons and $\mu_N$ is the Bohr magneton. The single particle matrix elements of the total angular momentum operator j are denoted by $j_v$.

For a zero spin ground state and individual dipole excitation the photon scattering cross section integrated over single level and the full solid angle is given by [12]

$$\sigma_{\gamma\gamma}(\omega_i) = \frac{11.5}{\omega_i^2} \Gamma_i [MeVmb] \qquad (16)$$

and ground state γ-decay width for M1 transitions is

$$\Gamma_i = 3.86 \, \omega_i^3 B(M1) 10^{-3} \text{ eV} \qquad (17)$$

where the excitation energy $\omega_i$ are in MeV and the B(M1) in $\mu_N^2$.

Numerical calculations were performed for the $^{140}$Ce with the spherical Woods-Saxon single-particle scheme [13]. The pairing parameters were calculated in accordance with [14] The strength parameter of the spin-spin interactions $\chi_{\sigma\tau}$ have been taken from ref.[9]. The calculated values of the B(M1), the radioactive width, the elastic scattering cross section $\sigma_\gamma$ and the two quasi-particle structure of the $1^+$ states with large B(M1) value and corresponding experimental data are listed in Table I. The calculations show that the appearance of collective $1^+$ states with large B(M1) is connected with transitions between levels of spin-orbit doublet. As can be seen from Table I the theory predicts strong collective $1^+$ state with M1 radiation width Γ=38.4 eV corresponding to $B(M1)=20.3\mu_N^2$ at the energy of 7.89 MeV. In formation of this state main contribution comes from the particle-hole transitions between levels of spin-orbit doublet : $1g_{9/2}^p - 1g_{7/2}^p$ and $1h_{11/2}^n - 1h_{9/2}^n$. The main strength(80%) of the M1 transition sum rule [15] is concentrated in this single state. The calculations have shown that all low lying $1^+$ states up to an energy on the order of 7 MeV, are characterised by small values of the B(M1) strength. However, the theory indicates the presence of the weakly collectivised spin-vibrational $1^+$ state at energy 4.52 MeV with $B(M1)=1.4\mu_N^2$. The rather large value of B(M1) is ensured by the large value of the M1 transition matrix element of the type $1d_{5/2}^p - 1d_{3/2}^p$ and contribution corresponding two-quasi-particle proton states to the norm of the wave function is about 99%. Finally, we note that neutron-proton mixing is as a rule weak in low-lying $1^+$ states. Strongly mixed states turn out to be in the energy region of the collective state. The contribution of the $1^+$ states in the resonance region usually exceed main of the energy weight sum rule. The agreement in resonance energy between the theoretical prediction and the experiment is rather good. The calculation predicts substantially more dipole strength than is indicated by experiment . This "quenching" is long standing problem of spin-flip M1-resonances [1]. The contribution from two-particle-two-hole channel [16] and Δ-isobar-nucleon hole mechanism [17] is not sufficient to explain the measured quenching. Another possible reason for this discrepancy in $^{140}$Ce could be that the strength in this nucleus exhibit strong fragmentation in the wide region energy between 7 and 12 MeV and a part of it might have just escaped detection. This experimentally observed fragmentation of M1 strength may be attributed to quadrupole deformation . Therefore, in the considered nucleus an essential role should play reciprocal effects between deformation and spin mixing [9].Recently experimental data confirming the deformation of N=82 semi-magic nuclei have been published [18,19].We suggest that deformation plays an important role in the fragmentation of the M1 strength in $^{140}$Ce and the study of the resonance in deformed base will provide important insight. Work along these lines is in progress.

Table Captions

TABLE I.- The integral characteristics of states $I^\pi=1^+$ in $^{140}$Ce. States are listed with the largest values of B(M1)

| | Theory ($\chi_{\sigma\tau}$=15/A MeV) | | | | Experiment [4,5] | |
|---|---|---|---|---|---|---|
| $\omega$, MeV | B(M1), $\mu_N^2$ | $\sigma_\gamma$, MeVmb | $\Gamma_\gamma$, eV | Structure of states | $\omega$, MeV | B(M1), $\mu_N^2$ |
| 4.52 | 1.4 | 0.29 | 0.51 | 99% pp $2d_{5/2}$-$2d_{3/2}$ | | |
| 7.89 | 20.3 | 7.07 | 38.4 | 74% nn $1h_{11/2}$-$1h_{9/2}$ 24% pp $1g_{9/2}$-$1g_{7/2}$ | 7.95 | 7.5 |
| 8.71 | 0.4 | 0.16 | 1.04 | 9% nn $1h_{11/2}$-$1h_{9/2}$ 90% nn $1g_{9/2}$-$1g_{7/2}$ | 9.08 | 0.39 |
| 11.91 | 1.7 | 0.94 | 11.57 | 98% pp $1h_{11/2}$-$1h_{9/2}$ | | |